\documentclass[12pt]{iopart}

\usepackage[compress]{cite}
\usepackage{graphicx}
\usepackage{bm}

\begin{document}

\title[Volatilities  of  FPT and FRT  on a small-world  scale-free network]{Volatilities analysis of  first-passage time and first-return time  on a small-world  scale-free network}

\author{Junhao Peng $^{1,2}$ }
\address{1.School of Math and Information Science, Guangzhou University, Guangzhou 510006, China. \\
2.Key Laboratory of Mathematics and Interdisciplinary Sciences of Guangdong
Higher Education Institutes, Guangzhou University, Guangzhou 510006, China.
} \ead{\mailto{pengjh@gzhu.edu.cn}}

\begin{abstract}

In this paper, we study random walks on  a small-world  scale-free network, also called as pseudofractal scale-free web (PSFW), and analyze the volatilities of  first passage time (FPT) and first return time (FRT) by using the variance and the reduced moment as the measures. Note that the FRT and FPT are deeply affected by the starting or target site. We don't intend to enumerate all the possible cases and analyze them. We only study  the volatilities of FRT for a given hub (i.e., node with highest degree) and the volatilities of  the global FPT (GFPT) to a given hub, which is  the average of the FPTs for arriving at a given hub from any possible starting site  selected randomly according to the equilibrium distribution of the Markov chain. Firstly, we calculate exactly  the probability generating function of the GFPT and FRT based on the self-similar structure of the PSFW. Then,  we  calculate the probability distribution, the mean, the variance and reduced moment of the GFPT and FRT by using the generating functions as a tool.  Results show that:  the reduced moment of FRT  grows with the increasing of the network order $N$ and  tends to infinity while $N\rightarrow\infty$;  but for the reduced moments of GFPT,  it is almost a constant($\approx1.1605$) for large $N$.   Therefore,  on the PSFW of large size,  the FRT  has  huge fluctuations and the estimate provided by MFRT is  unreliable, whereas  the fluctuations of the GFPT  is much smaller and the estimate provided by its mean  is more reliable. The method we propose can also be used to analyze the  volatilities of FPT and FRT on other networks with self-similar structure, such as  $(u, v)$ flowers and recursive scale-free trees.
\end{abstract}

\maketitle

\section{Introduction}
\label{intro}

First passage time (FPT), which is the time it takes a random walker to reach a given site for the first time, and first return time (FRT), which is the time it takes  a random walker to return to the starting site for the first time,  are two important quantities  in the random walk literature~\cite{Weiss94, HaBe87,  Avraham_Havlin04, ChPe13}.
The importance lies in the fact that  many physical processes are controlled by first passage events~\cite{Redner07, MeyChe11, VoRed10, Condamin05, MeKl00, CondaBe07, BeChKl10, BenVo14}, and that FRT can model the time intervals  between two successive extreme events~\cite{EiKaBu07,MoDa09,LiuJi09,SaKa08, HadLue02},  such as  traffic jams in roads, the  floods, the droughts, and power blackouts in electrical power grid, etc~\cite{ReThRe97, KondVa06,BatGer02}. Both FPT and FRT  are random variables which can not be determined exactly and researchers can only try to find suitable quantities to estimate  them.  A first step consists in the analysis of the mean of the two random variables, the mean first-passage time (MFPT) and the mean first return time (MFRT). For the MFRT, it can be calculated from the stationary distribution directly. That is to say,  the MFRT of node $v$ is $2m/d_v$ on any finite connected network, where $m$ is the total numbers of edges of the network and $d_v$ is the degree of node $v$~\cite{LO93, NR04}.   For the MFPT,  general formula to calculate the MPFT between any two nodes in any finite networks was presented ~\cite{NR04};  exact results of the MFPT to some special nodes and the mean trap time (i.e., the  MFPTs to a given target  averaged over all the starting nodes) have obtained on different networks, such as  Sierpinski  gaskets~\cite{KoBa02, BeTuKo10}, Apollonian network~\cite{ZhGuXi09},  scale-free Koch networks~\cite{ZhGa11, ZhGaxie10},  deterministic recursive trees~\cite{CoMi10,  ZhZhGa10,  ZhLi11,  ZhLiLin11,   Agl08, ZhYu09, WuLiZhCh12,AgCasCatSar15, Peng14a, Peng14b, Peng14c} and some other deterministic  networks~\cite{AgBu09, AgBuMa10, MeAgBeVo12, Zhsh12}. 

However, the MFPT and the MFRT aren't always the good estimates of the FPT  and the  FRT. Whether the estimates provided by the MFPT and the MFRT  are reliable  is subject to the  volatilities of the two random variables. 
It is well known that the  variance  and the reduced moment are good measures of random variable's volatilities~\cite{HaRo08}. A low variance or reduced moment indicates that the values of the random variable tend to be very close to the mean and shows the mean is  a good estimate of the random variable. On the contrary,  a high  variance or reduced moment indicates that  the random variable are spread out over a wider range of values and shows the  estimate provided by its mean is unreliable.  In the past several years, valuable results for the variance of the FPT are also obtained on different networks, such as  some tree and comb structure~\cite{Redner07, KahnRed89, KoBl90}, T-fractal, Sierpinski gasket, hierarchical percolation model and etc\cite{ArAnKo88, HaRo08}. 
But for the variance and the reduced moment of FRT and  FPT on the networks with scale-free and  small-world effect, to the best of our knowledge, explicit results are seldom reported.

The small-world  scale-free network considered here  is  a deterministically growing network which was introduced in Ref~\cite{Dorogovtsev02}.
It looks very similar to a fractal, but it has an infinite dimension~\cite{Dorogovtsev02, RoHa07}.  Therefore, it is not a fractal but only parody of it, and is called pseudofractal scale-free web (PSFW). As anticipated, the PSFW is a scale-free network with small-world effect~\cite{Dorogovtsev02, zhang07}. Its diameter  increases logarithmically with its size $N_t$ and has a degree distribution  $P(k) \sim k^{-\gamma}$  with $\gamma = 1 + \log 3 / \log 2 \approx 2.585$. Remarkably, values of $\gamma \in (2,3)$ are often evidenced in real growing scale-free networks~\cite{Dorogovtsev02}. Therefore,  the PSFW has attracted lots of attentions  in the past several years and much  effort has been devoted to the study of its properties, such as degree distribution, degree correlation, clustering coefficient~\cite{Dorogovtsev02,zhang07b}, diameter~\cite{zhang07b}, average path length~\cite{zhang07}, the number of spanning trees~\cite{zhang10}, and eigenvalues~\cite{ZhangLin15}. As for random walks on the PSFW, the  MFRT for any node $v$ is $2m/d_v$; Zhang and etc~\cite{ZhQiZh09} obtained the recursive relation of the  MFPT from any starting node to the hub (i.e.,  node with highest degree)  and then gained the mean trap time to the hub by averaging the MFPTs over all the possible starting nodes. However, the method of Ref.~\cite{ZhQiZh09} can not be used to derive  the variance of the FPT  because one can not derive the variance by its mean and other method to derive the higher moments of FPT and FRT are seldom reported. Therefore, whether the estimates provided by the MFPT and the MFRT  are reliable  is still unresolved. In order to obtain the higher moments of FPT and FRT, one should derive the the probability generating functions( or the probability distributions) of FPT and FRT.

 In this paper, we study the  volatilities of  the FRT  and the FPT  on the PSFW by using the variance and the reduced moment as the measure. Note that the FRT and FPT are deeply affected by the source or target site. We don't intend to enumerate all the possible cases and analyze them. On the contrary, we only study  the volatilities of FRT for a given hub  and the volatilityies of  the global FPT (GFPT) to a given hub, which is  the average of the FPTs for arriving at a given hub from any possible starting site  selected randomly according to the equilibrium distribution of the Markov chain (i.e., the probability that the random walker starts from  node $v$ is $d_v/(2m)$)~\cite{HvKa12}.
  Firstly, we calculate exactly  the probability generating function of the GFPT and FRT based on the self-similar structure of the PSFW. Then, exploiting the probability generating function tool, we  obtain the probability distribution, the mean, the variance and the reduced moment of the GFPT and FRT. Results show that:  the reduced moment of FRT  grows with the increasing of the network order $N$ and  it tends to infinity while $N\rightarrow\infty$;  but for the reduced moments of GFPT,  it is almost a constant($\approx1.1605$) on the PSFW of large size.
  Therefore,  on the PSFW of large size,  the FRT  has  huge fluctuations and the estimate provided by MFRT is  unreliable, whereas  the fluctuations of the GFPT  is much smaller and the estimate provided by its mean  is more reliable.

 This article is organized as follows. Section 2 presents the network model of the PSFW. Section 3  proposes a method to calculate  explicitly the probability generating function of the FRT and GFPT. Section 4 derives the exact  results for the probability distribution of the FRT and GFPT.   Section 5 analyzes the volatilities of  the GFPT. Section 6 analyzes the volatilities of  the FRT by using the variance and the reduced moment as the measures. Finally, Section 7 contains conclusions and discussions. Technical and lengthy calculations are collected in the Appendices.

\section{Network model}
\label{sec:psfw}
\begin{figure}
\begin{center}
\includegraphics[scale=0.6]{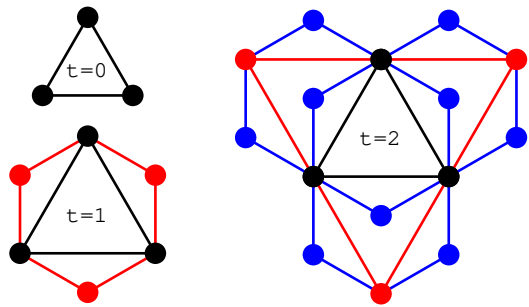}
\caption{The construction of the PSFW with generation $t=0, 1, 2$. At any iteration, each edge is replaced by a triangle.}
\label{fig:1}       
\end{center}
\end{figure}

The small-world  scale-free network considered here, also called as pseudofractal scale-free web (PSFW),  is  a deterministically growing network which can be  constructed iteratively~\cite{Dorogovtsev02}.  Let $G(t)$ denote the PSFW of generation $t$  ($t\geq 0$).   For $t=0$, $G(0)$ is a triangle. For $t\geq 1$, $G(t)$ is obtained  by replacing every edges of $G(t-1)$ with a triangle.  That is to say, for any edge of $G(t-1)$, a new node and two new edges  are introduced.  Figure \ref{fig:1} shows the construction of the PSFW of generation $t=0$, $1$, $2$. It is easy to see that the total number of edges for $G(t)$ is $E_t = 3^{t+1}$ and the total number of nodes for $G(t)$ is  $N_t=(3^{t+1}+3)/2$~\cite{Dorogovtsev02,zhang07b}.
By construction, at any iteration, the degree of the existing nodes is doubled. Therefore the three original nodes of the starting triangle have the highest degree and are called as the hubs of the PSFW in this paper.

\begin{figure}
\begin{center}
\includegraphics[scale=0.7]{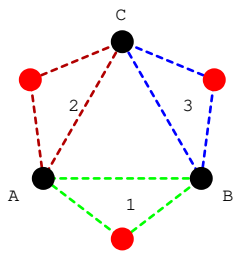}
\caption{Alternative construction of the PSFW that highlights self-similarity: the network of generation $t$, denoted by $G(t)$, is obtained by adjoining three copies of $G(t-1)$ at the hubs, denoted by $A$, $B$ and $C$.
}
\label{Self_similar}       
\end{center}
\end{figure}

The network also has an equivalent  construction method which highlights its self-similarity~\cite{zhang10, RoHa07}.
%
Referring to Figure \ref{Self_similar}, in order to obtain $G(t)$, one can make three copies of $G(t-1)$ and join them  at their hubs denoted by $A, B, C$.  In such a way, $G(t)$ is composed of three copies of $G(t-1)$ labeled as $\Gamma_1$, $\Gamma_2$, $\Gamma_3$, which are connected with each other by the three hubs. 
It is easy to know that the three hubs are equivalent and have the same properties. 

%

\section{Probability generating function of FRT and GFPT}
\label{PGF_RT_GFPT}

In this section we derive  exactly  the probability generating functions of the FRT for a given  hub and the GFPT to a given  hub. Not loss generality, we  only analyze the FRT for hub $C$ and GFPT to hub $C$ because the three hubs (i.e., $A$, $B$ and $C$) are equivalent and have the same properties. 

Let $F_{i\rightarrow j}(t,n)$ ($n=0$, $1$, $2$, $\cdots$)  be the probability distribution of FPT from node $i$ to hub $j$ (i.e., the probability that a random walker, starting at node $i$, first reaches the node $j$ at  time $n$ on network $G(t)$). Then $F_{C\rightarrow C}(t,n)$ ($n=0$, $1$, $2$, $\cdots$) is just the  probability distribution of FRT    for random walks starting from hub $C$ and  the  probability generating functions of FRT for hub $C$ is given by~\cite{NR04, Gut05}
\begin{equation}\label{PGF_FRT}
  \Phi_{FRT}(t,z)=\sum_{n=0}^{+\infty}z^nF_{C\rightarrow C}(t,n),
\end{equation}
where $t$ represents the generation of the PSFW.

 Note that, in the steady state~\cite{LO93}, the probability of finding the random walker at node $v$ is given by $d_v/2E_t$, where $d_v$ is the degree of node $v$. Averaged over all starting node $v$, the probability distribution of the GFPT to hub $C$  is defined by 
\begin{equation}\label{Def_GFPP}
  F_{C}(t, n)=\sum_{v}\frac{v}{2E_t}F_{v\rightarrow C}(t, n),
\end{equation}
where the sum runs over all the nodes of $G(t)$.
Then the probability generating functions of  GFPT to hub $C$ is represented as
\begin{equation}\label{PGF_GFPT}
  \Phi_{GFPT}(t,z)=\sum_{n=0}^{+\infty}z^nF_{C}(t, n).
\end{equation}
Results in Ref.~\cite{HvKa12} shows, $\Phi_{FRT}(t,z)$ and $\Phi_{GFPT}(t,z)$ can be calculated by

\begin{equation}\label{R_FRT_RT}
 \Phi_{FRT}(t,z)=1-1/ \Phi_{RT}(t,z)
\end{equation}
and
\begin{equation}\label{R_GFPT_RT}
  \Phi_{GFPT}(t,z)=\frac{z\cdot2^t}{(1-z)3^{t+1}}\times\frac{1}{ \Phi_{RT}(t,z)}
\end{equation}
respectively. Here $\Phi_{RT}(t,z)$  is the probability generating function of return time for hub $C$, which is defined by
  \begin{equation}\label{PGF_RTO}
 \Phi_{RT}(t,z)=\sum_{n=0}^{+\infty}z^nP(T_{C\rightarrow C}(t)=n),
\end{equation}
where $T_{C\rightarrow C}(t)$ denotes the return time and $P(T_{C\rightarrow C}(t)=n)$ is the  probability that a random walker, starting from hub $C$, is found at hub $C$ at time $n$ on network $G(t)$.

In order to calculate $\Phi_{FRT}(t,z)$ and $\Phi_{GFPT}(t,z)$  from Eqs.~(\ref{R_FRT_RT}) and (\ref{R_GFPT_RT}), we must calculate $\Phi_{RT}(t,z)$ firstly. Considering any return path $\pi$ of hub $C$ (i.e., starting from $C$ and ending at $C$). Its length is just the return time  $T_{C\rightarrow C}(t)$.  Let  $v_i$ be the node of $G(t)$ reached at time $i$. Then the path can be rewritten as $$\pi=(v_0=C,v_1,v_2\cdots,v_{T_{C\rightarrow C}(t)}=C).$$
In the meantime,  we denote with $\Omega$ the set of nodes $\{A,B, C \}$ and introduce the observable $\tau_i=\tau_i(\pi)$,
which is defined recursively as follows:
\begin{eqnarray}
\tau_0(\pi) &=& 0,\\
\tau_i(\pi) &=&\min\{k: k>\tau_{i-1},v_{\tau_i}\in \Omega, v_{\tau_i}\neq v_{\tau_{i-1}} \}.
\end{eqnarray}
Then,  considering only nodes in the set $\Omega$, the path $\pi$ can be restated into a ``simplified path'' defined as
\begin{equation}
\label{Def_simp}
  \iota(\pi)=(v_{\tau_{0}}=C,v_{\tau_1},\cdots, v_{\tau_L}=C ),
\end{equation}
where  $L=\max\{i: v_{\tau_i}=C\}$, which is the total number of observables obtained from the path $\pi$.
Therefore the path $\pi$ can be rewritten as
\begin{equation}
\label{ReDe_pi}
  \pi=\pi_{v_{\tau_{0}}\rightarrow v_{\tau_{1}}} || \pi_{v_{\tau_{1}}\rightarrow v_{\tau_{2}}} || \cdots || \pi_{v_{\tau_{L-1}}\rightarrow v_{\tau_{L}}} || \pi^{\overline{A},\overline{B}}_{C\rightarrow C} ,
\end{equation}
where $\pi_{v_{\tau_{i-1}}\rightarrow v_{\tau_{i}}}$ represents the sub-path from  $v_{\tau_{i-1}}$ to $ v_{\tau_{i}}$, $\pi^{\overline{A},\overline{B}}_{C\rightarrow C}$ represents the remaining sub-path after node $v_{\tau_{L}}$ and $\pi_{*} || \pi_{\diamond}$ is the concatenation of path $\pi_{*}$ and $\pi_{\diamond}$. Because the path  $\pi$ need not be a first return path of hub $C$ and maybe $v_{\tau_L}$ is not the last node of path  $\pi$. Therefore, $\pi^{\overline{A},\overline{B}}_{C\rightarrow C}$, which denotes the  remaining sub-path after node $v_{\tau_{L}}$,  is the return path from $C$ to $C$ which do not reach any new hubs (i.e., $A$ and $B$).

Let $T^{\overline{A},\overline{B}}_{C\rightarrow C}(t)$ denote the path length of $\pi^{\overline{A},\overline{B}}_{C\rightarrow C}$ and $T_i$ ($i=1,2,\cdots, L$) be the path length of $\pi_{v_{\tau_{i-1}}\rightarrow v_{\tau_{i}}}$, namely $T_i=\tau_i-\tau_{i-1}$. Therefore the  length of path  $\pi$ (i.e., the return time of hub $C$ on $G(t)$, which denoted as $T_{C\rightarrow C}(t)$)  satisfies
\begin{eqnarray}
\label{pathlength_1}
  T_{C\rightarrow C}(t)&=& T_1+T_2+\cdots+T_L+T^{\overline{A},\overline{B}}_{C\rightarrow C}(t).
\end{eqnarray}

 Note that the sub-path $\pi^{\overline{A},\overline{B}}_{C\rightarrow C}$ is the return path from $C$ to $C$ which may reach any node of $G(t)$ except $A$ and $B$. Therefore, it can be regarded as a return path of $C$ in the presence of two absorbing hubs (i.e., $A$ and $B$) on $G(t)$. Referring to Fig.~\ref{Self_similar},  we can also  find that, the sub-path $\pi^{\overline{A},\overline{B}}_{C\rightarrow C}$ only includes nodes of  $\Gamma_2$ and $\Gamma_3$, which are  copies of $G(t-1)$. According to the symmetric structure of the PSFW,  nodes of $\Gamma_3$ are in one to one corresponding with nodes of $\Gamma_2$. If we replaced all the nodes of  $\Gamma_3$ with the corresponding nodes of $\Gamma_2$ in the sub-path, we obtain a return path of $C$ in $\Gamma_2$ which never reaches hub $A$. It is a return path of $C$ in the presence of an absorbing hub  $A$ on $G(t-1)$ and has the same path length as $\pi^{\overline{A},\overline{B}}_{C\rightarrow C}$.  Therefore the  length of $\pi^{\overline{A},\overline{B}}_{C\rightarrow C}$ can be regarded as the return time of hub $C$ in the presence of an absorbing hub  $A$ on $G(t-1)$ and we denote it by $T^a_{C\rightarrow C}(t-1)$. Thus, Eq.~(\ref{pathlength_1}) can be rewritten as
\begin{eqnarray}
\label{pathlength}
  T_{C\rightarrow C}(t) &=& T_1+T_2+\cdots+T_L+T^a_{C\rightarrow C}(t-1).
\end{eqnarray}

Because the path $\iota(\pi)$ includes only the three nodes (i.e., $A$, $B$, $C$), which are just all the nodes of the PSFW with generation $0$, $\iota(\pi)$ is just a return path of $C$  on the PSFW of generation $0$ and $L$ is just the return time of hub $C$ on the PSFW of generation $0$. Therefore, the probability generating function of $L$ is $ \Phi_{RT}(0,z)$.


Note that $v_{\tau_i}\in \{A,B, C \}$ ($i=1,2,\cdots$, $L$) and nodes $A,B, C$ are hubs of  $\Gamma_1$, $\Gamma_2$, $\Gamma_3$, which are  copies of the PSFW of generation $t-1$. Then $T_i$ ($i=1,2,\cdots$, $L$) are identically distributed random variables, each of them can be regarded as the first-passage time from one hub to  another hub on the PSFW of generation $t-1$. For example, not loss generality,  assuming $v_{\tau_{i-1}}=A$ and $v_{\tau_i}=B$ for certain $i$. 
Because node  $B$ or $C$ does not appear between $v_{\tau_{i-1}}$  and $v_{\tau_{i-1}}$ in path  $\pi$,
  the sub-path from $v_{\tau_{i-1}}$ to $v_{\tau_{i}}$ only includes nodes of  $\Gamma_1$ and $\Gamma_2$ (see Fig.~\ref{Self_similar}) except $C$. According to the symmetric structure of the PSFW,  nodes of $\Gamma_1$ are in one to one corresponding with nodes of $\Gamma_2$. If we replaced all the nodes of  $\Gamma_2$ with the corresponding nodes of $\Gamma_1$ in the sub-path, we obtain a first-passage path of from $A$ to $B$ in $\Gamma_1$, , which is a  copy of $G(t-1)$. It has the same path length as the original sub-path. Therefore $T_i$ can be regarded as the first passage time from $A$ to $B$ on  $G(t-1)$.

Let $ \Phi_{FPT}(t,z)$   denote the  probability generating function of FPT from $A$ to $B$   and $ \Phi^a_{RT}(t,z)$ be the  probability generating function of the return time for hub $C$ in the presence of an absorbing hub  $A$ on $G(t)$. Therefore the probability generating function of  $T^a_{C\rightarrow C}(t-1)$ is $\Phi^a_{RT}(t-1,z)$ and  the probability generating functions of $T_i$ $(i=1,2, \cdots)$ is $\Phi_{FPT}(t-1,z)$.
 Note that  $L$, $T^a_{C\rightarrow C}(t-1)$, $T_1$, $T_2$, $\cdots$  are independent random variables.  According to the properties (see Eqs.~(\ref{Sum_PGF2}) and (\ref{Sum_PGFn})) of the  probability generating function presented in~\ref{sec:PGF}, the probability generating function of return time $T_{C\rightarrow C}(t)$  satisfies
\begin{equation}\label{RR_PGF_RTO}
  \Phi_{RT}(t,z)=\left. \Phi_{RT}(0,x)\right|_{x=\Phi_{FPT}(t-1,z)}*\Phi^a_{RT}(t-1,z).
\end{equation}
As for $\Phi_{RT}(0,x)$, $\Phi_{FPT}(t-1,z)$ and $\Phi^a_{RT}(t-1,z)$, we have calculated them in~\ref{PGF_g0}, \ref{PGF_FPTt} and \ref{PGF_aRTt} (see Eqs.~(\ref{PhiRT0}), (\ref{PhiFPTt}) and (\ref{PhiaRTt})). Therefore,
\begin{eqnarray}\label{Factor_RT}
  &&\left. \Phi_{RT}(0,x)\right|_{x=\Phi_{FPT}(t-1,z)}\nonumber \\
  &=&\Phi_{RT}(0,\frac{z}{2^{t} - z(2^{t}-1)})\nonumber \\
                 &=&\frac{\frac{z}{2^t -z(2^t-1)}-2}{[\frac{z}{2^t -z(2^t-1)}+2][\frac{z}{2^t -z(2^t-1)}-1] }                \nonumber \\
                 &=&\frac{[1-(1-\frac{1}{2^{t+1}})z][1-(1-\frac{1}{2^{t}})z]}{(1-z)[1-(1-\frac{3}{2^{t+1}})z]}.
\end{eqnarray}
Replacing $\left. \Phi_{RT}(0,x)\right|_{x=\Phi_{FPT}(t-1,z)}$ and $\Phi^a_{RT}(t-1,z)$ from Eqs.~(\ref{Factor_RT}) and (\ref{PhiaRTt}) respectively, we get

\begin{equation}\label{PGF_RTt}
  \Phi_{RT}(t,z)=\frac{[1-(1-\frac{1}{2^{t+1}})z]\prod_{i=1}^{t-1}[1-(1-\frac{1}{2^{i}})z]}{(1-z)\prod_{i=1}^{t+1}[1-(1-\frac{3}{2^{i}})z]},
\end{equation}
for any $t>0$. Inserting Eq.~(\ref{PGF_RTt}) into Eqs.~(\ref{R_FRT_RT}) and (\ref{R_GFPT_RT}), we obtain the rigorous solutions of $\Phi_{FRT}(t,z)$ and $\Phi_{GFPT}(t,z)$.

\section{Probability distribution of GFPT and FRT}
\label{P_FRT_GFPT}

\subsection{Probability distribution of GFPT to hub $C$}
\label{P_GFPT}
In this subsection, we calculate the  probability distribution of GFPT  by expanding the probability generating function $\Phi_{GFPT}(t,z)$ into a power series of $z$ and collecting the coefficient of $z^n$.
For the PSFW of generation $t=0$, replacing $\Phi_{RT}(0,z)$ from Eq.~(\ref{PhiRT0}) in Eqs.~(\ref{R_GFPT_RT}), we get
\begin{eqnarray}\label{Expand_PGF_GFPT0}
 \Phi_{GFPT}(0,z)&=&(z^2+2z)/(6-3z)    \nonumber \\
    &=&\frac{z^2+2z}{6}\times\frac{1}{1-\frac{z}{2}}         \nonumber \\
     &=&\frac{z^2+2z}{6}\sum_{n=0}^{+\infty}z^n\frac{1}{2^n} \nonumber \\
     &=&\frac{z}{3}+\sum_{n=2}^{+\infty}[\frac{2}{3}\times\frac{1}{2^{n-1}}z^n].
\end{eqnarray}
Therefore the  probability distribution of GFPT  on $G(0)$ read as  
 \begin{equation}\label{FPP_C_g0}
 F_{C}(0,n)=\left\{ \begin{array}{ll} 1/3 & {n=1} \\ \frac{2}{3}\times\frac{1}{2^{n-1}} & n\geq 2 \end{array} \right..
 \end{equation}

For the PSFW of generation $t\geq 1$, inserting Eq.~(\ref{PGF_RTt}) into Eq.~(\ref{R_GFPT_RT}), we obtain
the probability generating function of GFPT to hub $C$, i.e.,
 \begin{eqnarray}\label{PGF_GFPT}
   \Phi_{GFPT}(t,z)
     &=&\frac{2^tz}{3^{t+1}}\frac{\prod_{i=1}^{t+1}[1-(1-\frac{3}{2^{i}})z]}{[1-(1-\frac{1}{2^{t+1}})z]\prod_{i=1}^{t-1}[1-(1-\frac{1}{2^{i}})z]}\nonumber \\
     &=&\frac{2^t}{3^{t+1}}z(1+\frac{z}{2})(1-\frac{z}{4})\lambda(t,z)\nonumber \\
     &=&\frac{2^t}{3^{t+1}}(z+\frac{z^2}{4}-\frac{z^3}{8})\lambda(t,z),
\end{eqnarray}
where
\begin{eqnarray}\label{lamda}
\lambda(t,z)&=&\frac{\prod_{i=3}^{t+1}[1-(1-\frac{3}{2^{i}})z]}{[1-(1-\frac{1}{2^{t+1}})z]\prod_{i=1}^{t-1}[1-(1-\frac{1}{2^{i}})z]}\nonumber \\
     &=&\sum_{i=1}^{t-1}\frac{a_i}{1-(1-\frac{1}{2^{i}})z}+\frac{a_{t+1}}{1-(1-\frac{1}{2^{t+1}})z}.
\end{eqnarray}
Here $a_i=\left.\left\{[1-(1-\frac{1}{2^{i}})z]\lambda(t,z)\right\}\right|_{z=2^i/(2^i-1)}$ ($i=1$, $2$, $\cdots$, $t-1$, $t+1$).

Note that $\frac{1}{1-cz}=\sum_{n=0}^{+\infty}[{c^n}z^n]$. Expanding $\lambda(t,z)$ into a  power series of $z$,  inserting it into Eq.~(\ref{PGF_GFPT})  and calculating the coefficient of $z^n$,  we obtain the probability distribution of GFPT on $G(t)$, i.e.,

 for $n=1$,  $F_{C}(t,n)=\frac{2^t}{3^{t+1}}$;

  for $n=2$,  $F_{C}(t,n)=\frac{2^t}{3^{t+1}}\{\frac{1}{4}+a_{t+1}(1-1/2^{t+1})+\sum_{i=1}^{t-1}[a_i(1-1/2^i)]\}$;

 for $n\geq 3$,
\begin{equation}\label{GFPT_n}
\hspace{-15mm}F_{C}(t,n)=\frac{2^t}{3^{t+1}}\{a_{t+1}\mu_{t+1}(1-1/2^{t+1})^n+\sum_{i=1}^{t-1}[a_i\mu_i(1-1/2^i)^n]\}\propto(1-1/2^{t+1})^n,
\end{equation}
where 
\begin{equation}\label{mu_i}
\mu_i\equiv-\frac{1}{8}(1-1/2^{i})^{-3}+\frac{1}{4}(1-1/2^{i})^{-2}+(1-1/2^{i})^{-1},
\end{equation}
for $i=1$, $2$, $\cdots$, $t+1$.

The result shows that the global first passage probability $F_{C}(t,n)$ decreases exponentially with the number of steps $n$ while $n\rightarrow\infty$. We can also find that $F_{C}(t,n)\rightarrow 0$ while $t\rightarrow \infty$ for any  $n$, which  shows that the  global first passage probability $F_{C}(t,n)$  tends to zero with the increasing of the network for any fixed $n$.

\subsection{Probability distribution of FRT for hub $C$}
\label{P_FRT}
Similar to subsection~\ref{P_GFPT},  we calculate the  probability distribution of FRT  by expanding the probability generating function $\Phi_{FRT}(t,z)$ into a power series of $z$ and collecting the coefficient of $z^n$ in this subsection.

For the PSFW with generation $t=0$, replacing $\Phi_{RT}(0,z)$ from Eq.~(\ref{PhiRT0}) in Eqs.~(\ref{R_FRT_RT}), we get
\begin{eqnarray}\label{Expand_PGF_FRT0}
 \Phi_{FRT}(0,z)&=&z^2/(2-z)  \nonumber \\
 &=&\frac{z^2}{2}\times\frac{1}{1-\frac{z}{2}}         \nonumber \\
     &=&\frac{z^2}{2}\sum_{n=0}^{+\infty}z^n\frac{1}{2^n}.
\end{eqnarray}
Therefore the  probability distribution of FRT for hub $C$ reads as
 \begin{equation}\label{FRP_C_g0}
 F_{C\rightarrow C}(0,n)=\left\{ \begin{array}{ll} 0 & {n<2} \\ \frac{1}{2^{n-1}} & n\geq 2 \end{array} \right..
 \end{equation}

For the PSFW with generation $t\geq 1$, inserting Eq.~(\ref{PGF_RTt}) into Eq.~(\ref{R_FRT_RT}), we obtain the probability generating function of FRT for hub $C$, i.e.,
 \begin{eqnarray}\label{PGF_FRT}
 \Phi_{FRT}(t,z)
     &=&1-\frac{(1-z)\prod_{i=1}^{t+1}[1-(1-\frac{3}{2^{i}})z]}{[1-(1-\frac{1}{2^{t+1}})z]\prod_{i=1}^{t-1}[1-(1-\frac{1}{2^{i}})z]}\nonumber \\
     &=&1-(1-z)(1+\frac{z}{2})(1-\frac{z}{4})\lambda(t,z)\nonumber \\
     &=&1-(1-\frac{3}{4}z-\frac{3}{8}z^2+\frac{1}{8}z^3)\lambda(t,z),
\end{eqnarray}
%
where $\lambda(t,z)$ is defined by Eq.~(\ref{lamda}).

By expanding $\lambda(t,z)$ into a  power series of $z$,  inserting it into Eq.~(\ref{PGF_FRT})  and calculating the coefficient of $z^n$,
we obtain the  probability distribution of FRT for hub $C$ on $G(t)$, i.e.,

 for $n=0$ and $n=1$,
  $F_{C\rightarrow C}(t,n)=0$;

   for $n=2$,  $F_{C\rightarrow C}(t,n)=\frac{1}{3}+\frac{2}{3}\times4^{-(t+1)}$;

for $n\geq 3$,
\begin{equation}\label{FRP_n}
\hspace{-10mm}F_{C\rightarrow C}(t,n)=a_{t+1}\omega_{t+1}(1-1/2^{t+1})^n+\sum_{i=1}^{t-1}[a_i\omega_i(1-1/2^i)^n]\propto (1-1/2^{t+1})^n,
\end{equation}
where $a_i$ ($i=1$, $2$, $\cdots$, $t-1$, $t+1$) are defined in subsection~\ref{P_GFPT} and for $i=1$, $2$, $\cdots$, $t+1$,
\begin{equation}\label{omega_i}
\omega_i\equiv-\frac{1}{8}(1-1/2^{i})^{-3}+\frac{3}{8}(1-1/2^{i})^{-2}+\frac{3}{4}(1-1/2^{i})^{-1}-1.
\end{equation}

Similar to the  global first passage probability, the first return probability $F_{C\rightarrow C}(t,n)$ also decreases exponentially with the number of steps $n$ while $n\rightarrow\infty$. However, we find $\mu_1=2$, $\omega_1 =1$ and $\mu_i\rightarrow \frac{9}{8}$,  $\omega_i\rightarrow 0$ while $(i\rightarrow+\infty)$. Therefore, $\mu_{t+1}\approx \mu_{1}$ and $\omega_{t+1}\ll \omega_{1}$ for big $t$. Thus the leading term (i.e., $ a_{t+1}\omega_{t+1}(1-1/2^{t+1})^n$) of Eq.~(\ref{FRP_n}) has no great influence on $F_{C\rightarrow C}(t,n)$ while $n$ is  small because $\omega_{t+1}\ll \omega_{1}$;  whereas the leading term (i.e., $ a_{t+1}\mu_{t+1}(1-1/2^{t+1})^n$) of Eq.~(20) has great influence on the global first passage probability $F_{C}(t,n)$ almost for any $n$ because $\mu_{t+1}\approx \mu_{1}$. 

We also find that, for small $n$, the first return probability $F_{C\rightarrow C}(t,n)$ is very high even on the PSFW  with large scale. For example  $F_{C\rightarrow C}(t,n)> \frac{1}{3}$  for $n=2$ and any $t>0$.
The results show that, the random walker has high probability of returning to hub $C$  in short time, but with increasing of the number of the steps $n$, the first return probability decreases exponentially with $n$. The reason is that, in a very short time  after the random walker leaves the starting position (i.e., the number of steps $n$ is very small), the random walker is  closed to the starting position and he has high probability of returning to origin; but with  time elapsed (i.e., the number of steps $n$ increased),  the position distribution of the random walker tends to the stationary distribution, the time the random walker first returns to the starting node is similar to the time he first reaches the starting node. Therefore, the first return probability shows similarly asymptotic behaviors with the global first passage probability while $n\rightarrow\infty$.

\section{Volatilities analysis of the GFPT}
\label{Sta_GFPT}
In this section, we first present the recurrence relations that the probability generating function $\Phi_{GFPT}(t,z)$, the first and second moment of the GFPT to hub $C$ satisfy, Then we derive exactly formulas for the  first and the second moment  of the GFPT. Finally we obtain  the variance and the reduced moment~\cite{HaRo08} of the GFPT, which are the measures  for the volatilities of the GFPT.

It is easy to obtain from Eq.~(\ref{PGF_RTt}) that 
\begin{eqnarray}\label{RR_PGF_RTt}
  \Phi_{RT}(t,z)&=&\Phi_{RT}(t\!-\!1,z)\times\frac{[1\!-\!(1\!-\!\frac{1}{2^{t\!-\!1}})z][1\!-\!(1\!-\!\frac{1}{2^{t\!+\!1}})z]}{[1\!-\!(1\!-\!\frac{1}{2^{t}})z][1\!-\!(1\!-\!\frac{3}{2^{t\!+\!1}})z]} \nonumber \\
  &\equiv&\Phi_{RT}(t\!-\!1,z)\times\frac{\theta(t,z)}{\rho(t,z)}
\end{eqnarray}
hold for any $t>0$. Here $\theta(t,z)\equiv[1\!-\!(1\!-\!\frac{1}{2^{t\!-\!1}})z][1\!-\!(1\!-\!\frac{1}{2^{t\!+\!1}})z]$ and $\rho(t,z)\equiv[1\!-\!(1\!-\!\frac{1}{2^{t}})z][1\!-\!(1\!-\!\frac{3}{2^{t\!+\!1}})z]$.

Inserting Eq.~(\ref{RR_PGF_RTt}) into Eq.~(\ref{R_GFPT_RT}), we find $\Phi_{GFPT}(t,z)$ satisfies the  following recurrence relation:
\begin{eqnarray}\label{RR_PGF_GFPT}
  \Phi_{GFPRT}(t,z)
  &=&\frac{2^t\cdot z}{3^{t+1}(1-z)}\times\frac{1}{\Phi_{RT}(t-1,z)}\times\frac{\theta(t,z)}{\rho(t,z)}\nonumber \\
  &=&\frac{2}{3}\frac{\rho(t,z)}{\theta(t,z)}\Phi_{GFPT}(t-1,z),
\end{eqnarray}
with initial condition
\begin{equation}\label{PGF_GFPT_g0}
 \Phi_{GFPT}(0,z)=\frac{z(z+2)}{3(2-z)}.
\end{equation}
By calculating the first order derivative with respect to $z$ on both sides of Eq.~(\ref{RR_PGF_GFPT}), we obtain
\begin{eqnarray}\label{First_d_GFPT}
  \hspace{-15mm}&&\frac{\partial}{\partial z}\Phi_{GFPT}(t,z)\nonumber \\
  \hspace{-15mm}&=&\frac{2}{3}\left\{\frac{\partial}{\partial z}\left(\frac{\rho(t,z)}{\theta(t,z)}\right)\times \Phi_{FRT}(t-1,z)+\frac{\rho(t,z)}{\theta(t,z)}\times\frac{\partial}{\partial z}\Phi_{FRT}(t-1,z)\right\}
  \nonumber \\
  \hspace{-15mm}&=& \frac{2}{3}\left\{\left[\frac{z^2(\frac{5}{4}8^{-t}-4^{-t})+z 4^{-t}}{[\theta(t,z)]^2}\right]\Phi_{FRT}(t-1,z)+\frac{\rho(t,z)}{\theta(t,z)}\times\frac{\partial}{\partial z}\Phi_{FRT}(t-1,z)\right\}
  .
\end{eqnarray}
Noting that $\Phi_{GFPT}(t,1)=1$, $\rho(t,1)=3\times 2^{-2t-1}$, $\theta(t,1)=2^{-2t}$ and $\left. \frac{\partial}{\partial z}\Phi_{GFPT}(0,z)\right|_{z=1}=\frac{7}{3}$ and letting  $z=1$ in Eq.~(\ref{First_d_GFPT}), we obtain  the first moment of GFPT to hub $C$, i.e.,
\begin{eqnarray}\label{First_m_GFPT}
 \langle GFPT(t) \rangle&=&\left. \frac{\partial}{\partial z}\Phi_{GFPT}(t,z)\right|_{z=1}\nonumber \\
 &=& \langle GFPT(t-1) \rangle+\frac{5}{6}\times 2^t  \nonumber \\
  &=& \langle GFPT(0) \rangle+\frac{5}{6}\times \prod_{k=1}^t{2^k}\nonumber \\
  &=& \frac{5}{3}\times{2^{t}}+\frac{2}{3},
\end{eqnarray}
where  $\langle GFPT(t) \rangle$ denotes the mean of GFPT to hub $C$ on $G(t)$. The result is  consistent with the  asymptotic behaviors of the   probability distribution of GFPT as shown in subsection~\ref{P_GFPT}.

Similarity, we can also obtain the second moment of the GFPT to hub $C$, referred to as $\langle GFPT^2 (t)\rangle$.
By taking the first order derivative on both sides of Eq.~(\ref{First_d_GFPT}), we obtain
   \begin{eqnarray}\label{Second_d_GFPT}
     \hspace{-18mm} &&\frac{\partial^2}{\partial z^2}\Phi_{GFPT}(t,z)\nonumber \\
      \hspace{-18mm}&=& \frac{2}{3}\left\{\frac{\rho(t,z)}{\theta(t,z)}\times\frac{\partial^2}{\partial z^2}\Phi_{GFPT}(t-1,z)     +2\left[\frac{z^2(\frac{5}{4}8^{-t}-4^{-t})+z 4^{-t}}{[\theta(t,z)]^2}\right]\frac{\partial}{\partial z}\Phi_{GFPT}(t-1,z)\right.\nonumber \\
     \hspace{-18mm}  & &+\frac{\Phi_{FRT}(t-1,z)-1}{[\theta(t,z)]^4} \left[[2z(\frac{5}{4}8^{-t}-4^{-t})+ 4^{-t}][\theta(t,z)]^2\right.  \nonumber \\
     \hspace{-18mm} & &\left.\left.2[z^2(\frac{5}{4}8^{-t}-4^{-t})+z 4^{-t}]\times\theta(t,z)\times\frac{\partial}{\partial z}\theta(t,z)\right]\right\}.
  \end{eqnarray}

Note that $\Phi_{FRT}(t,1)=1$, $\rho(t,1)=3\times 2^{-2t-1}$, $\theta(t,1)=2^{-2t}$ and $ \left.\frac{\partial}{\partial z}{\theta(t,z)}\right|_{z=1}=2^{1-2t}-5\times 2^{-t-1}$. 
By posing $z=1$ in  Eq.~(\ref{Second_d_GFPT}), we obtain
   \begin{eqnarray}
     \left. \frac{\partial^2}{\partial z^2}\Phi_{GFPT}(t,z)\right|_{z=1}=\left.\frac{\partial^2}{\partial z^2}\Phi_{GFPT}(t-1,z)\right|_{z=1}+\frac{44}{9}\times4^t-\frac{5}{9}\times2^t.
  \end{eqnarray}
  Therefore, the second moment of GFPT of node $C$  satisfies
  \begin{eqnarray}\label{R_Second_m_GFPT}
      \langle GFPT^2(t)\rangle &=&\left. \frac{\partial^2}{\partial z^2}\Phi_{GFPT}(t,z)\right|_{z=1}+\langle GFPT(t) \rangle \nonumber\\
       &=&\langle GFPT^2(t-1)\rangle+\frac{44}{9}\times4^t+\frac{5}{18}\times2^t.
  \end{eqnarray}
 As for the initial value $\langle GFPT^2(0)\rangle$, it can be obtained by calculating the first and second order derivative with respect to $z$ on both sides of Eq.~(\ref{PGF_GFPT_g0}) and fixing $z=1$, i.e.,
  \begin{eqnarray}\label{Second_m_g0_GFPT}
      \langle GFPT^2(0)\rangle=23/3.
  \end{eqnarray}
Using Eq.~(\ref{R_Second_m_GFPT}) recursively, we get
  \begin{eqnarray}\label{Second_m_gt_GFPT}
     \langle GFPT^2(t)\rangle
     &=&\langle GFPT^2(t-1)\rangle+\frac{44}{9}\times4^t+\frac{5}{18}\times2^t\nonumber\\
     &=&\cdots \nonumber\\
     &=&\langle GFPT^2(0)\rangle+\frac{44}{9}\times\prod_{k=0}^{t}4^k+\frac{5}{18}\times\prod_{k=0}^{t}2^k\nonumber\\
     &=&\frac{176}{27}\times4^t+\frac{5}{9}\times2^t+\frac{16}{27}\nonumber\\
     &\sim&N_t^{2ln2/ln3}.
  \end{eqnarray}

Therefore the  variance of GFPT to node $C$  turns out to be 
  \begin{eqnarray}\label{Var}
     Var(GFPT(t))
     &=&\langle GFPT^2(t)\rangle-[\langle GFPT(t)\rangle]^2\nonumber\\
     &=&\frac{101}{27}\times4^t-\frac{5}{3}\times2^t+\frac{4}{27}\nonumber\\
     &\approx & \frac{101}{27}\times\langle GFPT(t)\rangle^2
     .
  \end{eqnarray}
Therefore the  variance of GFPT to node $C$ scales quadratically with its mean value, 
and the reduced moment $R(GFPT(t))=\sqrt{Var(GFPT(t))} / \langle GFPT(t)\rangle$~\cite{HaRo08}
is almost a constant. In the limit of large size,
  \begin{eqnarray}\label{Reduce_m_FRT}
     \lim\limits_{t \to \infty }{R(GFPT(t))}
     =\sqrt{{101}/{75}}\approx1.1605.
  \end{eqnarray}
%
\section{Volatilities analysis of FRT }
\label{Sta_FRT}
In this section, we first present the recurrence relations that the probability generating function $\Phi_{FRT}(t,z)$, the first and second moment of the FRT for hub $C$ satisfy, Then we derive exactly formulas for the  first and the second moment  of the FRT for hub $C$. Finally we obtain  the variance and the reduced moment~\cite{HaRo08} of the FRT, which are the measures for the volatilities of the FRT.


Note that Eq.~(\ref{R_FRT_RT}) can be rewriten as
\begin{equation}\label{R_RT_FRT}
 \Phi_{RT}(t,z)=1/ [1-\Phi_{FRT}(t,z)].
\end{equation}
Therefore $\Phi_{FRT}(t,z)$ satisfies the  following recurrence relation:
\begin{eqnarray}\label{RR_PGF_FRTt}
  \Phi_{FRT}(t,z)&=&1-1/\Phi_{RT}(t,z)\nonumber \\
  &=&1-1/[\Phi_{RT}(t\!-\!1,z)\times\frac{\theta(t,z)}{\rho(t,z)}]\nonumber \\
  &=&1-1/[\frac{1}{[1-\Phi_{FRT}(t-1,z)]}\times\frac{\theta(t,z)}{\rho(t,z)}] \nonumber \\
  &=&\frac{\rho(t,z)}{\theta(t,z)}[\Phi_{FRT}(t-1,z)-1]+1,
\end{eqnarray}
with initial condition
 $\Phi_{FRT}(0,z)=1-1/ \Phi_{RT}(0,z)=z^2/(2-z)$.
Calculating the first order derivative with respect to $z$ on both sides of Eq.~(\ref{RR_PGF_FRTt}), we obtain
\begin{eqnarray}\label{First_d}
  \hspace{-23mm}\frac{\partial}{\partial z}\Phi_{FRT}(t,z)=\frac{\partial}{\partial z}\left(\frac{\rho(t,z)}{\theta(t,z)}\right)\times [\Phi_{FRT}(t-1,z)-1]+\frac{\rho(t,z)}{\theta(t,z)}\times\frac{\partial}{\partial z}\Phi_{FRT}(t-1,z)
  .
\end{eqnarray}
Noting that $\Phi_{FRT}(t,1)=1$, $\rho(t,1)=3\times 2^{-2t-1}$, $\theta(t,1)=2^{-2t}$ and $\left. \frac{\partial}{\partial z}\Phi_{FRT}(0,z)\right|_{z=1}={3}$ and letting  $z=1$ in Eq.~(\ref{First_d}), we obtain  the first moment of FRT of hub $C$, i.e.,
\begin{eqnarray}\label{First_m}
 \langle FRT(t) \rangle&=&\left. \frac{\partial}{\partial z}\Phi_{FRT}(t,z)\right|_{z=1}= \frac{3}{2}\times  \langle FRT(t-1)  \rangle\nonumber \\
  &=& \frac{3^t}{2^t} \langle FRT(0)  \rangle=\frac{3^{t+1}}{2^t},
\end{eqnarray}
where  $ \langle FRT(t) \rangle$ denotes the mean first return time of hub $C$ on $G(t)$. The result is  consistent with the result calculated by stationary distribution~\cite{LO93}. Differing from the result  of the mean of GFPT, the mean of the FRT does not scale with $t$ as $2^t$, but scale with $t$ as $(\frac{3}{2})^t$. It is not consistent with the  asymptotic behaviors of the   probability distribution of FRT as shown in subsection~\ref{P_FRT}. The reason is that the leading term (i.e., $ a_{t+1}\omega_{t+1}(1-1/2^{t+1})^n$) of Eq.~(\ref{FRP_n}), which shows the asymptotic behavior,  has no great influence on the first return probability $F_{C\rightarrow C}(t,n)$ while $n$ is small.
 The asymptotic behavior of the first return probability just shows its property while $n$ is big enough.  In fact, the random walker has high probability of returning to origin in short time which can not be shown by the asymptotic behaviors of the  probability distribution. 

Similarity, we can also obtain the second moment of the FTR of hub $C$, referred to as $\langle FRT^2 (t)\rangle$.
By taking the first order derivative on both sides of Eq.~(\ref{First_d}), we obtain
   \begin{eqnarray}\label{Second_d}
     &&\frac{\partial^2}{\partial z^2}\Phi_{FRT}(t,z)\nonumber \\
     &=&\frac{\rho(t,z)}{\theta(t,z)}\times\frac{\partial^2}{\partial z^2}\Phi_{FRT}(t-1,z)+2\frac{\partial}{\partial z}\left(\frac{\rho(t,z)}{\theta(t,z)}\right)\times\frac{\partial}{\partial z}\Phi_{FRT}(t-1,z)\nonumber \\
  &+&{[\Phi_{FRT}(t-1,z)-1]}\frac{\partial^2}{\partial z^2}\left(\frac{\rho(t,z)}{\theta(t,z)}\right).
  \end{eqnarray}
Note that $\Phi_{FRT}(t,1)=1$, $\rho(t,1)=3\times 2^{-2t-1}$, $\theta(t,1)=2^{-2t}$, $\left. \frac{\partial}{\partial z}\Phi_{FRT}(t,z)\right|_{z=1}=\frac{3^{t+1}}{2^t}$ and
\begin{eqnarray}
 \frac{\partial}{\partial z}\left(\frac{\rho(t,z)}{\theta(t,z)}\right)=\frac{z^2(\frac{5}{4}8^{-t}-4^{-t})+z 4^{-t}}{[\theta(t,z)]^2}.
\end{eqnarray}
By posing $z=1$ in  Eq.~(\ref{Second_d}), we obtain
 \begin{eqnarray}\label{T2_Ck}
      \left. \frac{\partial^2}{\partial z^2}\Phi_{FRT}(t,z)\right|_{z=1}=\frac{3}{2}\times\left.\frac{\partial^2}{\partial z^2}\Phi_{FRT}(t-1,z)\right|_{z=1}+5\times3^t.
  \end{eqnarray}
  Therefore, the second moment of FRT of node $C$  satisfies
  \begin{eqnarray}\label{R_Second_m}
      & &\langle FRT^2(t)\rangle\nonumber\\
      &=&\left. \frac{\partial^2}{\partial z^2}\Phi_{FRT}(t,z)\right|_{z=1}+\langle FRT(t) \rangle \nonumber\\
      &=&\frac{3}{2}\left[\left. \frac{\partial^2}{\partial z^2}\Phi_{FRT}(t-1,z)\right|_{z=1}+\langle FRT(t-1) \rangle\right]+5\times3^t \nonumber\\
     &=&\frac{3}{2}\langle FRT^2(t-1)\rangle+5\times3^t \nonumber\\
     &=&\cdots      \nonumber\\
     &=&\frac{3^t}{2^t}\langle FRT^2(0)\rangle+5\times3^t\sum_{k=0}^{t-1}2^{-k}.
  \end{eqnarray}
As for $\langle FRT^2(0)\rangle$, it  can be obtained by calculating the first and second order derivative with respect to $z$ on both sides of Eq.~(\ref{PhiRT0}) and fixing $z=1$, i.e.,
  \begin{eqnarray}\label{Second_m_g0}
      \langle FRT^2(0)\rangle=11.
  \end{eqnarray}
Inserting Eq.~(\ref{Second_m_g0}) into Eq.~(\ref{R_Second_m}), we have
  \begin{eqnarray}\label{Second_m_gt}
     \langle FRT^2(t)\rangle=\frac{3^t}{2^t}+10\times3^t.
  \end{eqnarray}
Since the volume of the underlying structure scales as $N_t \sim 3^t$ for large sizes, the previous expression can be restated as
\begin{equation}\label{Scale_2M}
\langle FRT^2(t)\rangle \sim N_t,
\end{equation}
namely the second moment of FRT scales linearly with the volume $N_t$. We can also obtain
 the  variance of FRT which is shown as 
  \begin{eqnarray}\label{Var}
     Var(FRT(t))
     &=&\langle FRT^2(t)\rangle-[\langle FRT(t)\rangle]^2\nonumber\\
     &=&3^t\left[10+\frac{1}{2^t}-9\times\frac{3^{t}}{4^{t}}\right] \nonumber\\
     &\approx & \frac{10}{9}\times(\frac{4}{3})^t\times\langle FRT(t) \rangle^2.
  \end{eqnarray}
Hence  the reduced moment, defined by $R(FRT(t))=\sqrt{Var(FRT(t))} / \langle FRT(t)\rangle$~\cite{HaRo08}, grows with the increasing of the network and in the limit of large size,
  \begin{eqnarray}\label{Reduce_m_FRT}
     \lim\limits_{t \to \infty }{R(FRT(t))}
     =\lim\limits_{t \to \infty }\frac{\sqrt{10}}{3}\times \left(\frac{4}{3}\right)^{t/2}=\infty.
  \end{eqnarray}
 Therefore, on the PSFW of large size, the FRT of node $C$ has  huge fluctuations and the estimate provided by MFRT is unreliable.
 By comparing  the  result with that of the GFPT, we find the fluctuations of the GFPT is much smaller  and the estimate of the GFPT provided by its mean is more reliable.

\section{Conclusions}
\label{sec:4}

We have analyzed the volatilities of  the FRT  and the GFPT  on the PSFW by using the variance  as the measure. Results show that: on the PSFW of large size, the FRT of a given hub has  huge fluctuation and the estimate provided by MFRT is  unreliable, whereas  the fluctuation of the GFPT to a given hub is much smaller and the estimate provided by its mean is more reliable. Whether it is always the case on other networks is an interesting problem unresolved. Of course, the method proposed here  can also be used on other self-similar graph such as $(u, v)$ flower, T-graph, recursive  scale-free trees and etc.

\ack{
The author is grateful to E. Agliari and the anonymous referees for their valuable  suggestions.
This work was supported  by the scientific research program of Guangzhou municipal colleges and universities under Grant No. 2012A022.
}

\appendix
 \section{Probability generating function and its properties}
\label{sec:PGF}
  Let $T$ be a discrete random variable which takes only non-negative integer values, and whose probability distribution is $p_k$ ($k=0,1,2,\cdots$) . The probability generating function of $T$ is defined as 
 \begin{equation}\label{Def_PGF}
   \Phi_{T}(z)=\sum_{k=0}^{+\infty}z^k p_k.
 \end{equation}
 The probability generating function of $T$ is  determined by the probability distribution and, in turn, it uniquely determines the probability distribution. If $T_1$ and $T_2$ are two random variables with the same probability generating function, then they have the same probability distribution. Given the probability generating function $\Phi_{T}(z)$ of the random variable $T$, we can obtain the probability distribution $p_k$ ($k=0,1,2,\cdots$)  as the coefficient of $z^k$ in the Taylor's series expansion of $\Phi_{T}(z)$ about $z=0$. 

Also, the $n$-th moment $\langle T^n \rangle \equiv \sum_{k=0}^{+\infty} k^n p_k$, can be written in terms of combinations of derivatives (up to the $n$-th order) of $\Phi_{T}(z)$ calculated in $z=1$. In particular,
\begin{eqnarray}\label{n_moment}
\langle T \rangle  &=& \frac{\partial \Phi_{T}(z)}{ \partial z} \Big |_{z=1},\\
\langle T^2 \rangle  &=&  \frac{\partial^2 \Phi_{T}(z)}{ \partial z^2} \Big|_{z=1} + \frac{\partial \Phi_{T}(z)}{ \partial z} \Big|_{z=1}.
 \end{eqnarray}

Finally, we list  some useful properties of the probability generating function~\cite{Gut05}, which would be useful in this paper:
   \begin{itemize}
  \item Let $T_1$ and $T_2$ be two independent random variables with probability generating functions $\Phi_{T_1}(z)$ and $\Phi_{T_2}(z)$, respectively. Then, the probability generating function of random variable $T_1+T_2$ reads as
  \begin{equation}\label{Sum_PGF2}
    \Phi_{T_1+T_2}(z)=\Phi_{T_1}(z)\Phi_{T_2}(z).
 \end{equation}
  \item Let $N$, $T_1$, $T_2$, $\cdots$ be independent random variables. If $T_i$ ($i=1, 2, \cdots$) are identically distributed, each with probability generating function $\Phi_{T}(z)$, and, being $\Phi_{N}(z)$ the probability generating function of $N$, the random variable defined as
 \begin{equation}\label{Sum_R_V}
    S_N=T_1+T_2+\cdots+T_N
 \end{equation}
 has probability generating function
  \begin{equation}\label{Sum_PGFn}
    \Phi_{S_N}(z)=\Phi_{N}(\Phi_{T}(z)).
 \end{equation}
 \end{itemize}

\section{PGF of FPT and RT on the PSFW of generation $0$  }
\label{PGF_g0}
Let
 $$M=(P_{xy})_{3\times3}$$
  be the  transition probability matrix for random walks on the PSFW of generation $0$. This means
\begin{equation}
\label{Pxy}
  P_{xy}=\left\{ \begin{array}{ll}
  \frac{1}{d_x} & \textrm{if $x\sim y$, and $x$ is not an absorbing node}\\
   0 & \textrm{others} \end{array} \right.,
\end{equation}
where $x\sim y$ means that there is an edge between $x$ and $y$ and $d_x$ is the degree of node $x$.
Then we can calculate the probability generating function  directly by
 \begin{eqnarray}\label{Formular_PGF}
\Psi(z)=\sum_{n=0}^{+\infty}(zM)^n=(I-zM)^{-1},
\end{eqnarray}
where $\Psi(z)=(\psi_{xy}(z))_{3\times3}$ and  $\psi_{xy}(z)$ is the probability generating function of passage time from node $x$ to $y$, whereas $\psi_{xx}(z)$ is the probability generating function of return time of node  $x$.  If $y$ is an  absorbing node, $\psi_{xy}(z)$ is just the probability generating function of first passage time from node $x$ to $y$.

\paragraph*{Exact calculation of  $\Phi_{RT}(0,z)$.}
In this case, no node is absorbing node. Therefore,
\begin{equation}
\label{PM1}
  M=\left( \begin{array}{lll}
  0 & \frac{1}{2} & \frac{1}{2}  \\
  \frac{1}{2} & 0 & \frac{1}{2} \\
   \frac{1}{2} & \frac{1}{2} & 0
   \end{array} \right).
\end{equation}
Inserting Eq.~(\ref{PM1}) into Eq.~(\ref{Formular_PGF}), we obtain $\Psi(z)$ for this case. (Note: as a matter of fact, we use \emph{Matlab} for this task.) Then the probability generating function of the RT of hub $C$  is
\begin{equation}\label{PhiRT0}
 \Phi_{RT}(0,z)=\psi_{33}(z)=(z-2)/[(z-1)(z+2)].
\end{equation}

\paragraph*{Exact calculation of  $\Phi_{FPT}(0,z)$ and $\Phi^a_{RT}(0,z)$.}
Let  hub $A$  be the absorbing node.  Then all coordinates of the first row of the  transition probability matrix $M$ are assigned $0$. Therefore,
\begin{equation}
\label{PM2}
  M=\left( \begin{array}{lll}
  0 & 0 & 0 \\
   \frac{1}{2} & 0 & \frac{1}{2} \\
     \frac{1}{2} & \frac{1}{2} &0
   \end{array} \right).
\end{equation}
 Calculating $\Psi(z)$ from Eq.~(\ref{Formular_PGF}) by using \emph{Matlab},  the probability generating function of the FPT from $B$ to $A$ is
\begin{equation}\label{PhiFPT0}
 \Phi_{FPT}(0,z)= \psi_{21}(z)=\frac{-z}{z-2},
\end{equation}
and the probability generating function of the return time of hub $C$ in the presence of an absorbing hub $A$ reads as
\begin{equation}\label{PhiaRT0}
  \Phi_{RT}^a(0,z)= \psi_{33}(z)=\frac{-4}{(z-2)(z+2)}.
\end{equation}

\section{Probability generating function of FPT from $A$ to $B$}
 \label{PGF_FPTt}
For the PSFW of generation $t=0$, the probability generating function of  FPT from $A$ to $B$ is presented in Eq.~(\ref{PhiFPT0}). 
For the PSFW of generation $t>0$, let $T_{A\rightarrow B}(t)$ denote the FPT from $A$ to $B$. Similar to the derivation of Eq.~(\ref{pathlength}), we can find independent random variables $L$,  $T_1$, $T_2$, $\cdots$, such that
\begin{equation}
\label{FPT_pathlength}
  T_{A\rightarrow B}(t)= T_1+T_2+\cdots+T_L.
\end{equation}
Here $L$ is the first-passage time from hub $A$ to $B$  on $G(0)$ and $T_i$ ($i=1,2,\cdots$) are identically distributed random variables, each of them is the first-passage time from one hub to  another  hub on $G(t-1)$. Therefore the probability generating function of $L$ is $\Phi_{FPT}(0,z)$ and the probability generating function of $T_i$ ($i=1,2,\cdots$) is $\Phi_{FPT}(t-1,z)$. Thus, we can obtain from Eqs.~(\ref{Sum_PGF2}), (\ref{Sum_PGFn}) and (\ref{FPT_pathlength}) that the probability generating function of $T_{A\rightarrow B}(t)$ satisfies
\begin{equation}\label{Rec_PGF_FPT}
  \Phi_{FPT}(t,z)=\Phi_{FPT}(0,\Phi_{FPT}(t-1,z)).
\end{equation}
Therefore, for any $t\geq0$,
\begin{equation}\label{PhiFPTt}
  \Phi_{FPT}(t,z)=\frac{z}{2^{t+1} - z(2^{t+1}-1)},
\end{equation}
which is proved by  mathematical induction as follows.

For $t=0$, we find  Eq.~(\ref{PhiFPTt}) holds from Eq.~(\ref{PhiFPT0}). Assuming that
Eq.~(\ref{PhiFPTt}) holds for  $t=k-1$ $(k\geq 1)$, we will prove it also holds for $t=k$. In fact,
\begin{eqnarray}
   \Phi_{FPT}(k,z)&=& \Phi_{FPT}(0, \Phi_{FPT}(k-1,z))\nonumber \\
                 &=& \Phi_{FPT}(0,\frac{z}{2^k-z(2^k-1)})\nonumber \\
                 &=&\frac{\frac{z}{z(2^k-1)-2^k}}{\frac{-z}{z(2^k-1)-2^k}-2 }                \nonumber \\
                 &=&\frac{z}{2^{k+1} - z(2^{k+1}-1)}.
\end{eqnarray}
Therefore, Eq.~(\ref{PhiFPTt}) holds for  $t=k$, and this ends the proof.

\section{Probability generating function of return time for hub $C$ in the presence of an absorbing hub $A$}
 \label{PGF_aRTt}
For the PSFW of generation $t=0$, the probability generating function of return time for hub $C$ is presented in Eq.~(\ref{PhiaRT0}).
For the PSFW of generation $t>0$, let $T^a_{C\rightarrow C}(t)$ denote the return time of $C$ in the presence of an absorbing hub $A$  on $G(t)$. Similar to the derivation of Eq.~(\ref{pathlength}), we can find independent random variables $L$, $T^a_{C\rightarrow C}(t-1)$, $T_1$, $T_2$, $\cdots$,  such that
\begin{equation}
\label{RTa_pathlength}
  T^a_{C\rightarrow C}(t)= T_1+T_2+\cdots+T_L+ T^a_{C\rightarrow C}(t-1).
\end{equation}
Here $L$ is the return time of hub $C$ in the presence of absorbing hub $A$  on $G(0)$ and $T_i$ ($i=1,2,\cdots$) are identically distributed random variables, each of them is the first-passage time from one hub to  another  hub on $G(t-1)$. Therefore the probability generating function of $L$ is $\Phi^a_{RT}(0,z)$ and the probability generating function of $T_i$ ($i=1,2,\cdots$) is $\Phi_{FPT}(t-1,z)$. Thus, we can obtain from Eqs.~(\ref{Sum_PGF2}), (\ref{Sum_PGFn}) and (\ref{RTa_pathlength}) that the probability generating function of $T^a_{A\rightarrow B}(t)$ satisfies
\begin{equation}\label{Rec_PGF_FPT}  
  \Phi^a_{RT}(t,z)=\Phi^a_{RT}(0,\Phi_{FPT}(t-1,z))*\Phi^a_{RT}(t-1,z).
\end{equation}
Using Eq.~(\ref{Rec_PGF_FPT}) recursively,
 \begin{eqnarray}\label{PhiaRTt1}
  \Phi^a_{RT}(t,z)&=&\Phi^a_{RT}(0,z)\prod_{k=0}^{t-1}\Phi^a_{RT}(0,\Phi_{FPT}(k,z))
\end{eqnarray}
Recalling Eqs.~(\ref{PhiaRT0}) and (\ref{PhiFPTt}), for any $k\geq 0$,
\begin{eqnarray}\label{Factor_RR}   
    && \Phi^a_{RT}(0,\Phi_{FPT}(k,z)) \nonumber \\
    &=&\Phi^a_{RT}(0,\frac{z}{2^{k+1} - z(2^{k+1}-1)})\nonumber \\
    &=&\frac{-4}{\left[\frac{z}{2^{k+1} -z(2^{k+1}-1)}-2\right]\left[(\frac{z}{2^{k+1} -z(2^{k+1}-1)}+2\right] }                \nonumber \\
      &=&\frac{\left[1-(1-1/2^{k+1})z\right]^2}{\left[1-(1-1/2^{k+2})z\right]\left[1-(1-3/2^{k+2})z\right]}.
\end{eqnarray}
 Replacing $\Phi^a_{RT}(0,\Phi_{FPT}(k,z))$ from Eq.~(\ref{Factor_RR}), we obtain
 \begin{eqnarray}\label{PhiaRTt}
  \Phi^a_{RT}(t,z)&=&\Phi^a_{RT}(0,z)\prod_{k=0}^{t-1}\Phi^a_{RT}(0,\Phi_{FPT}(k,z))\nonumber \\
                 &=&\frac{\prod_{k=1}^{t}[1-(1-\frac{1}{2^{k}})z]}{[1-(1-\frac{1}{2^{t+1}})z]\prod_{k=1}^{t+1}[1-(1-\frac{3}{2^{k}})z] }.
\end{eqnarray}


\end{document}